\documentclass[runningheads]{llncs}
\usepackage{tikz}
\usetikzlibrary{arrows.meta, positioning, fit, backgrounds}
\usepackage{bm}

\usepackage[T1]{fontenc}
\usepackage{amsmath}
\usepackage{amssymb}
\usepackage{microtype}
\usepackage{graphicx}
\usepackage{url}
\usepackage{eso-pic}
\AddToShipoutPictureBG*{\AtPageUpperLeft{\put(\LenToUnit{0.5\paperwidth},\LenToUnit{-1.3cm}){\makebox(100pt,0)[c]{\parbox{\paperwidth}{\small\itshape Accepted Extended Abstract, 7th International Workshop on Active Inference (IWAI 2026),14--16 October 2026, Madrid, Spain}}}}}
\begin{document}
\title{Framing Spatial Hearing within the Free Energy Principle: A Hierarchical Model of Perceptual Learning}
\titlerunning{Spatial Hearing within the Free Energy Principle}
%
\author{Roberto Barumerli\inst{1} \and Nils Meyer-Kahlen\inst{1,2} \and Lorenzo Picinali\inst{1}}
\authorrunning{Barumerli et al.}
%
\institute{Dyson School of Design Engineering, Imperial College London, UK
 \and
Dept. of Information and Communications Engineering, Aalto University, Finland
\email{r.barumerli@imperial.ac.uk}}
\maketitle              
%

\begin{abstract}
Spatial hearing relies on dynamic internal representations to decode spatial information from ever-changing acoustic conditions. A fundamental example is sound localisation, where listeners infer the direction of an invisible sound. Bayesian inference provides a computational account of such a localisation process; however, modelling efforts are largely limited to static listening scenarios. Instead, the free energy principle (FEP) describes perception, learning, and action as joint minimisation of variational free energy (VFE), providing a principled framework to account for dynamic conditions. Here, we take a first step toward applying the FEP to spatial hearing. We characterise how listeners' beliefs evolve across trials, in which they estimated sound direction and received subsequent visual feedback on each trial.
Listeners are described by a hierarchical generative model where visual feedback drives gradient descent on VFE, updating beliefs over three independent quantities: lateral and polar localisation precisions and hemifield disambiguation. The model captures individual learning trajectories and reveals components with different dynamics: lateral precision improves, as does hemifield disambiguation, consistent with increased sensitivity of the auditory pathway in decoding vertical direction. Notably, polar precision also increases, most for listeners with the lowest initial precision, suggesting that the observed vertical improvement is driven by both hemifield disambiguation and polar precision. This work establishes spatial hearing as a tractable domain for FEP and motivates future closed-loop experiments with full action-perception cycles.
\end{abstract}
\noindent\textbf{Introduction.} Understanding how the auditory system decodes spatial information is a central challenge in hearing research, yet computational accounts largely focus on static scenarios~\cite{majdak_amt_2022,barumerli2023}, leaving its dynamic, experience-dependent nature under-explored~\cite{barumerli2025}. Auditory spatial perception is inherently plastic~\cite{van_der_heijden_cortical_2019} and localisation involves both perception and action~\cite{middlebrooks2015}, making the FEP, a unified computational account of perception, learning, and action~\cite{friston2010,parr2022active}, a natural framework.
Localisation improves across trials with visual feedback~\cite{majdak2010sound}, yet no mechanistic account exists of how individual beliefs evolve trial by trial.
Here, we use the FEP~\cite{bogacz_tutorial_2017} to model perceptual learning as belief revision, decomposing localisation improvement into three components with distinct timescales, validated on behavioural data from~\cite{majdak2010sound}.

\begin{figure}[t]
  \centering
  \includegraphics[width=.7\textwidth]{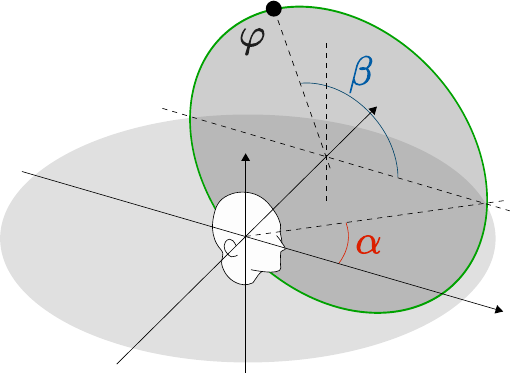}
  \caption{Interaural-polar coordinate system~\cite{morimoto1984}. The source direction is defined by $\boldsymbol{\varphi} = (\alpha, \beta)$: lateral angle $\alpha$ identifies the interaural axis and the cone of confusion (green circle); polar angle $\beta$ defines elevation within that cone. Reference points: front $(0^\circ,\,0^\circ)$, back $(0^\circ,\,180^\circ)$, sides $(\pm90^\circ,\,0^\circ)$.}
  \label{fig:reference}
\end{figure}

\noindent\textbf{Problem formulation.} On each trial $i$, the listener estimates the direction of an invisible sound source $\boldsymbol{\varphi}_t$ and produces a pointing response $\boldsymbol{\varphi}_r$, followed by visual feedback at $\boldsymbol{\varphi}_t$. Directions are expressed in interaural-polar coordinates $\boldsymbol{\varphi} = (\alpha, \beta)$ ~\cite{morimoto1984}, separating the contributions of auditory cues to behavioural error (Fig.~\ref{fig:reference}): the lateral angle $\alpha$, informed by interaural time and level differences, defines the cone of confusion, within which the polar angle $\beta$ is decoded from spectral-shape cues~\cite{middlebrooks2015}. With uniform priors on $\boldsymbol{\varphi}_r$ and $\boldsymbol{\varphi}_t$, the listener model factorises as $
p(\boldsymbol{\varphi}_r|\boldsymbol{\varphi}_t, \boldsymbol{\theta}) = p(\alpha_r|\alpha_t, \boldsymbol{\theta})\,p(\beta_r|\alpha_r, \beta_t, \boldsymbol{\theta})$ where $\boldsymbol{\theta}$ captures the listener's beliefs about lateral and polar precisions, and hemifield disambiguation (i.e., front--back confusions~\cite{middlebrooks2015}).

We employ a parametric formulation to preserve differentiability within FEP. The lateral component is a von Mises distribution with location $\alpha_t$ and concentration $\kappa_L$ (i.e., lateral precision). The polar component is a two-component von Mises mixture with concentration $\kappa_P$ (i.e., polar precision), locations $\beta_t$ and $\pi-\beta_t$ (i.e., correct and incorrect hemifields) and hemifield weight $w$; $\kappa_P$ is scaled by $\cos^2(\alpha_r)$, reflecting the reduced relevance of spectral cues at lateral extremes, where the interaural cone collapses (Fig.~\ref{fig:reference})~\cite{morimoto1984,mardia_directional_2009}.

\begin{figure}[t]
  \centering

\begin{tikzpicture}[
    > = {Stealth[length=5pt, width=4pt]},
    lvl2/.style = {
        draw = blue!55!black, fill = blue!7,
        rounded corners = 5pt, thick,
        text width = 5.6cm, minimum height = 1.7cm,
        align = center, inner sep = 7pt,
        font = \small
    },
    lvl1/.style = {
        draw = orange!70!black, fill = orange!7,
        rounded corners = 5pt, thick,
        text width = 5.6cm, minimum height = 1.7cm,
        align = center, inner sep = 7pt,
        font = \small
    },
    ext/.style = {
        font = \footnotesize,
        align = center,
        text width = 2.2cm
    },
    arr/.style  = {->, thick},
    darr/.style = {->, thick, dashed, gray!70}
]

\node[lvl2] (L2) {%
    \textbf{Level 2 --- Belief Revision}\\[3pt]
    Natural-gradient descent on $\mathcal{F}$\\[1pt]
    $\bm{\theta} = (\kappa_L,\; \kappa_P,\; w)$%
};

\node[lvl1, below = 1.5cm of L2] (L1) {%
    \textbf{Level 1 --- Perceptual Inference}\\[3pt]
    $p(\alpha_r \mid \alpha_t,\, \kappa_L)$: VM\\[1pt]
    $p(\beta_r \mid  \alpha_r,\, \beta_t,\, \kappa_P,\, w)$: VM mixture%
};

\draw[arr, blue!55!black]
    ([xshift = 7pt] L2.south)
    -- node[right, font=\scriptsize, text width=2cm, align=left,
            xshift=2pt] {beliefs\\$(\kappa_L\; \kappa_P,\; w)$}
    ([xshift = 7pt] L1.north);

\draw[arr, orange!70!black]
    ([xshift = -7pt] L1.north)
    -- node[left, font=\scriptsize, text width=1.8cm, align=right,
            xshift=-2pt] {prediction\\error $\nabla_{\bm{\theta}}\mathcal{F}$}
    ([xshift = -7pt] L2.south);

\node[ext, left = .5cm of L2] (vf) {%
    $(\kappa_{P,0},\, w_0,\, \kappa_{L,0})$%

};

\node[ext, left = .5cm of L1] (aud) {%
    (1) Sound\\presentation
    $\boldsymbol{\varphi}_t = (\alpha_t,\, \beta_t)$%
};

\node[ext, right = .5cm of L2] (init) {%
    (3) Visual feedback\\[2pt]
    $\boldsymbol{\varphi}_t$%
};

\node[ext, right = .5cm of L1] (resp) {%
    (2) Pointing response\\[2pt]
    $\boldsymbol{\varphi}_r = (\alpha_r,\, \beta_r)$%
};

\draw[arr] (vf.east)    -- (L2.west);
\draw[arr] (aud.east)   -- (L1.west);
\draw[arr] (L1.east)    -- (resp.west);
\draw[darr](init.west)  -- (L2.east);

\begin{scope}[on background layer]
    \node[
        draw = gray!55, dashed,
        rounded corners = 8pt,
        fill = gray!4,
        fit = (L2)(L1),
        inner xsep = 7pt,
        inner ysep = 7pt,
        label = {[font=\footnotesize\itshape, text=gray!70,
                  yshift=3pt] above: FEP Agent}
    ] {};
\end{scope}

\end{tikzpicture}
  \caption{Architecture of the hierarchical FEP agent. Level~1 encodes the
    listener's generative model of auditory space as a von Mises mixture (polar)
    and a single von Mises (lateral). Level~2 revises the agent's beliefs
    $\boldsymbol{\theta} = (\kappa_L, \kappa_P, w)$ via natural-gradient descent on variational
    free energy $\mathcal{F}$, using visual feedback $\boldsymbol{\varphi}_t = (\alpha_t, \beta_t)$ as a
    high-precision teaching signal. Initial beliefs are estimated from
    baseline block available from the original experiment (naive localisation without feedback~\cite{majdak2010sound}) by maximum likelihood. Learning
    rates $(\eta_L, \eta_P, \eta_w)$ are fitted per subject.}
  \label{fig:arch}
\end{figure}
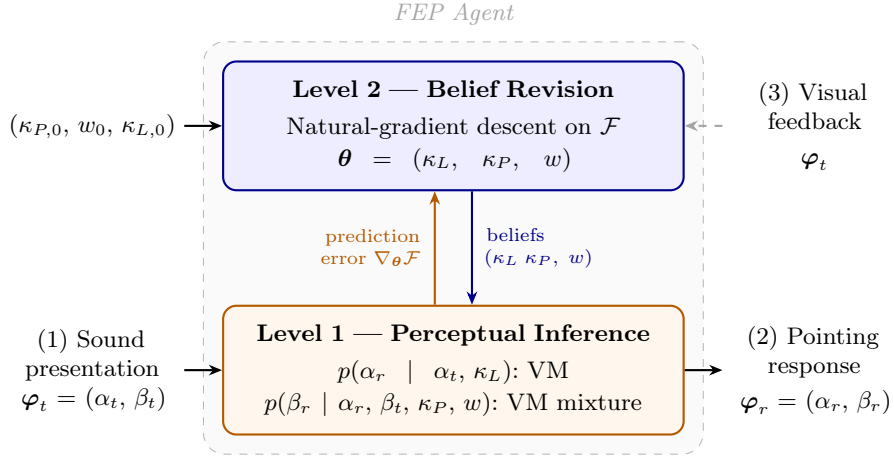

Belief revision follows from minimising variational free energy $\mathcal{F}$ with respect to $\boldsymbol{\theta} = (\kappa_L, \kappa_P, w)$. With a flat prior and Laplace approximation~\cite{buckley_free_2017}, $\mathcal{F}$ reduces to $-\log p(\bm{\varphi}_r|\bm{\varphi}_t, \bm{\theta})$, so belief revision amounts to online likelihood ascent: a deliberately minimal instance of the FEP, covering perception and learning but not action (Fig.~\ref{fig:arch}). Each belief takes a natural-gradient step~\cite{amari1998} after every trial, scaled by the inverse Fisher information to update components living on different scales consistently (problem formulation available in Appendix~\ref{app:formulation}). Updates are sequential, producing a trajectory that batch optimisation cannot recover.
Per subject (N = 5), learning rates $(\eta_L, \eta_P, \eta_w)$ were estimated by maximum likelihood~\cite{acerbi2017practical}, and $\boldsymbol{\theta}_0$ from the baseline block~\cite{majdak2010sound} (Appendix~\ref{app:parest}).

\begin{figure}[t]
  \centering
  \includegraphics[width=.95\textwidth]{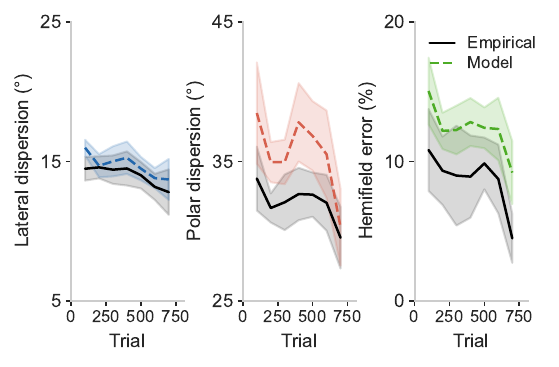}
  \includegraphics[width=.95\textwidth]{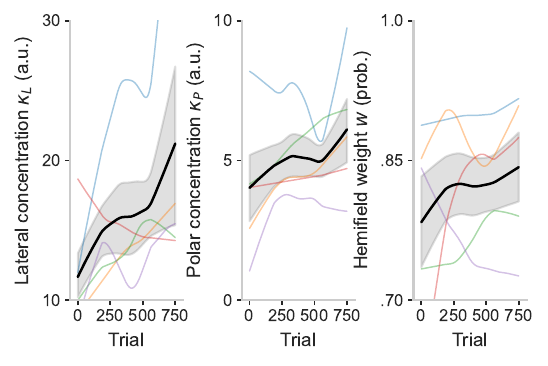}
  \caption{\emph{Top:} Group-average behavioural trajectories, manual condition ($n{=}5$). Solid: empirical (mean $\pm$\,1\,standard error (SE) ribbon); dashed: FEP model. Left to right:  lateral dispersion (standard deviation (SD) of lateral error, $^\circ$); polar dispersion (SD of polar error excluding hemifield errors, $\cos^2\alpha_t$-weighted, $^\circ$); hemifield error rate (proportion of responses deviating more than $90^\circ$ from the target, $\cos^2\alpha_t$-weighted, \%). The model captures the direction and shape of all three metrics.
  \emph{Bottom:} Individual parameter trajectories, manual group ($n{=}5$), truncated to 750 trials (LOESS-smoothed, span\,=\,0.5; thin coloured lines: individual subjects; thick black line: mean $\pm$\,1\,SE ribbon).}
  \label{fig:params}
\end{figure}

\noindent\textbf{Results and Discussion.} The model captures group-average trajectories ($R^2 = 0.75, 0.83, 0.79$ for lateral, polar and hemifield errors, Fig.~\ref{fig:params}, top). Individual parameter trajectories (Fig.~\ref{fig:params}, bottom) differ across components: $\kappa_L$ rises for most subjects; $\kappa_P$ rises, proportionally more for lower baseline $\kappa_P$; $w$ increases steadily (median $\eta_L, \eta_P, \eta_w = 0.010, 0.015, 0.014$; Appendix~\ref{sec:app_robust}).
These distinct trajectories are consistent with previously reported mechanisms, including sensorimotor recalibration~\cite{majdak2010sound}, stable spectral cue-to-direction mappings acquired through natural listening~\cite{carlile1997}, and increased sensitivity to altered spectral cues~\cite{hofman1998relearning}.
Notably, behavioural polar dispersion decreases and $\kappa_P$ rises under the $\cos^2$ weighting: the improvement seen in aggregate statistics stems from hemifield disambiguation and sharper polar precision (Fig.~\ref{fig:params}, middle column).

Our approach has limitations: a small cohort and a perception-only loop in which pointing is a read-out and feedback is external. Two extensions would cast spatial hearing as full active inference: informative priors on $\boldsymbol{\theta}$, yielding posterior uncertainty on $\kappa_P$, and head movements as policies, resolving hemifield ambiguity by action rather than by learning.

\bibliographystyle{splncs04}
\bibliography{references}

@article{hofman1998relearning,
  author    = {Hofman, Paul M. and Van Riswick, Jos G. A. and Van Opstal, A. John},
  title     = {Relearning sound localization with new ears},
  journal   = {Nature Neuroscience},
  volume    = {1},
  number    = {5},
  pages     = {417--421},
  year      = {1998},
  doi       = {10.1038/1633}
}

@article{majdak2010sound,
  author    = {Majdak, Piotr and Goupell, Matthew J. and Laback, Bernhard},
  title     = {3-{D} localization of virtual sound sources: Effects of visual
               environment, pointing method, and training},
  journal   = {Attention, Perception, \& Psychophysics},
  volume    = {72},
  number    = {2},
  pages     = {454--469},
  year      = {2010},
  doi       = {10.3758/APP.72.2.454}
}

@article{friston2010,
  author    = {Friston, Karl J.},
  title     = {The free-energy principle: a unified brain theory?},
  journal   = {Nature Reviews Neuroscience},
  volume    = {11},
  number    = {2},
  pages     = {127--138},
  year      = {2010},
  doi       = {10.1038/nrn2787}
}

@article{majdak_amt_2022,
	title = {{AMT} 1.x: {A} toolbox for reproducible research in auditory modeling},
	volume = {6},
	issn = {2681-4617},
	shorttitle = {{AMT} 1.x},
	doi = {10.1051/aacus/2022011},
	language = {en},
	urldate = {2022-08-12},
	journal = {Acta Acustica},
	author = {Majdak, Piotr and Hollomey, Clara and Baumgartner, Robert},
	year = {2022},
	pages = {19},
}

@article{amari1998,
  author    = {Amari, Shun-ichi},
  title     = {Natural Gradient Works Efficiently in Learning},
  journal   = {Neural Computation},
  volume    = {10},
  number    = {2},
  pages     = {251--276},
  year      = {1998},
  doi       = {10.1162/089976698300017746}
}

@article{carlile1997,
  author    = {Carlile, Simon and Leong, Patricia and Hyams, Sarah},
  title     = {The nature and distribution of errors in sound localization
               by human listeners},
  journal   = {Hearing Research},
  volume    = {114},
  number    = {1--2},
  pages     = {179--196},
  year      = {1997},
  doi       = {10.1016/S0378-5955(97)00161-5}
}

@article{bogacz_tutorial_2017,
	title = {A tutorial on the free-energy framework for modelling perception and learning},
	volume = {76},
	issn = {00222496},
	doi = {10.1016/j.jmp.2015.11.003},
	language = {en},
	urldate = {2023-01-16},
	journal = {Journal of Mathematical Psychology},
	author = {Bogacz, Rafal},
	month = feb,
	year = {2017},
	pages = {198--211},
}

@article{buckley_free_2017,
	title = {The free energy principle for action and perception: {A} mathematical review},
	volume = {81},
	issn = {00222496},
	shorttitle = {The free energy principle for action and perception},
	doi = {10.1016/j.jmp.2017.09.004},
	language = {en},
	urldate = {2026-05-29},
	journal = {Journal of Mathematical Psychology},
	author = {Buckley, Christopher L. and Kim, Chang Sub and McGregor, Simon and Seth, Anil K.},
	month = dec,
	year = {2017},
	pages = {55--79},
}

@book{parr2022active,
  author    = {Parr, Thomas and Pezzulo, Giovanni and Friston, Karl J.},
  title     = {Active Inference: The Free Energy Principle in Mind, Brain,
               and Behavior},
  publisher = {MIT Press},
  year      = {2022},
  isbn      = {978-0-262-36997-8},
  doi = {10.7551/mitpress/12441.001.0001}
}

@inproceedings{acerbi2017practical,
author = {Acerbi, Luigi and Ma, Wei Ji},
title = {Practical Bayesian optimization for model fitting with Bayesian adaptive direct search},
year = {2017},
isbn = {9781510860964},
publisher = {Curran Associates Inc.},
address = {Red Hook, NY, USA},
booktitle = {Proceedings of the 31st International Conference on Neural Information Processing Systems},
pages = {1834–1844},
numpages = {11},
location = {Long Beach, California, USA},
series = {NIPS'17}
}

@incollection{middlebrooks2015,
	title = {Sound localization},
	volume = {129},
	isbn = {978-0-444-62630-1},
	doi = {10.1016/B978-0-444-62630-1.00006-8},
	language = {en},
	urldate = {2020-02-26},
	booktitle = {Handbook of {Clinical} {Neurology}},
	publisher = {Elsevier},
	author = {Middlebrooks, John C.},
	year = {2015},
	pages = {99--116},
}

@article{van_der_heijden_cortical_2019,
	title = {Cortical mechanisms of spatial hearing},
	volume = {20},
	issn = {1471-003X, 1471-0048},
	doi = {10.1038/s41583-019-0206-5},
	language = {en},
	number = {10},
	urldate = {2020-02-13},
	journal = {Nature Reviews Neuroscience},
	author = {van der Heijden, Kiki and Rauschecker, Josef P. and de Gelder, Beatrice and Formisano, Elia},
	month = oct,
	year = {2019},
	pages = {609--623},
}

@article{morimoto1984,
	title = {Localization cues of sound sources in the upper hemisphere},
	volume = {5},
	doi = {10.1250/ast.5.165},
	number = {3},
	journal = {Journal of the Acoustical Society of Japan (E)},
	author = {Morimoto, Masayuki and Aokata, Hitoshi},
	year = {1984},
	pages = {165--173},
}

@book{mardia_directional_2009,
	title = {Directional statistics},
	publisher = {John Wiley \& Sons},
	author = {Mardia, Kanti V and Jupp, Peter E},
	year = {2009},
}

@article{barumerli2025,
	title = {{FrAMBI}: {A} {Software} {Framework} for {Auditory} {Modeling} {Based} on {Bayesian} {Inference}},
	volume = {23},
	issn = {1559-0089},
	shorttitle = {{FrAMBI}},
	doi = {10.1007/s12021-024-09702-5},
	language = {en},
	number = {2},
	urldate = {2025-02-18},
	journal = {Neuroinformatics},
	author = {Barumerli, Roberto and Majdak, Piotr},
	month = feb,
	year = {2025},
	pages = {20},
}

@article{barumerli2023,
	title = {A {Bayesian} model for human directional localization of broadband static sound sources},
	volume = {7},
	copyright = {© The Author(s), Published by EDP Sciences, 2023},
	issn = {2681-4617},
	doi = {10.1051/aacus/2023006},
	language = {en},
	urldate = {2026-02-13},
	journal = {Acta Acustica},
	publisher = {EDP Sciences},
	author = {Barumerli, Roberto and Majdak, Piotr and Geronazzo, Michele and Meijer, David and Avanzini, Federico and Baumgartner, Robert},
	year = {2023},
	pages = {12},
}


\clearpage

\clearpage
\setcounter{equation}{0}
\renewcommand{\theequation}{A\arabic{equation}}
\setcounter{section}{0}
\renewcommand{\thesection}{\Alph{section}}
\renewcommand{\thesubsection}{\thesection.\arabic{subsection}}

\section{Appendix}

Code and data are available here: \\\url{https://github.com/robaru/iwai2026-fep-spatial-hearing}.

\subsection{Model formulation}
\label{app:formulation}
\subsubsection{Generative model}
Directions are expressed in interaural-polar coordinates $\boldsymbol{\varphi} = (\alpha, \beta)$~\cite{morimoto1984}. Circular quantities are modelled with the von Mises distribution
\begin{equation}
\mathcal{VM}(x;\,\mu,\kappa) = \frac{e^{\kappa\cos(x-\mu)}}{2\pi I_0(\kappa)},
\end{equation}
where $I_0$ is the modified Bessel function of order zero and $\kappa$ the concentration (precision). The listener model factorises into a lateral and a polar component, $p(\boldsymbol{\varphi}_r|\boldsymbol{\varphi}_t, \boldsymbol{\theta}) = p(\alpha_r|\alpha_t, \kappa_L)\,p(\beta_r|\alpha_r, \beta_t, \kappa_P, w)$, with
\begin{align}
p(\alpha_r|\alpha_t, \kappa_L) &= \mathcal{VM}(\alpha_r;\,\alpha_t,\kappa_L),\\
p(\beta_r|\alpha_r, \beta_t, \kappa_P, w) &= w\,\mathcal{VM}(\beta_r;\,\beta_t,\tilde\kappa_P) + (1-w)\,\mathcal{VM}(\beta_r;\,\pi-\beta_t,\tilde\kappa_P),
\end{align}
where $\tilde\kappa_P = \kappa_P\cos^2(\alpha_r)$ accounts for the geometric compression of the cone of confusion at lateral angles, and $w \in (0,1)$ is the probability of resolving the correct hemifield. The two mixture components share the same concentration: hemifield confusion affects where the response lands, not how tightly it clusters.

\subsubsection{Free energy and natural gradient}
For a recognition density $q(\boldsymbol{\theta})$ over the beliefs $\boldsymbol{\theta} = (\kappa_L, \kappa_P, w)$, the variational free energy at trial $i$ is
\begin{equation}
\mathcal{F}_i = \mathrm{KL}\!\left[q(\boldsymbol{\theta})\,\|\,p(\boldsymbol{\theta})\right] - \mathbb{E}_{q}\!\left[\log p(\boldsymbol{\varphi}_r|\boldsymbol{\varphi}_t, \boldsymbol{\theta})\right].
\label{eq:app_F}
\end{equation}
Under a Laplace approximation ($q$ Gaussian centred on the current point estimate $\boldsymbol{\theta}_i$) and a flat prior, the complexity term is constant in $\boldsymbol{\theta}_i$, so $\mathcal{F}_i = -\log p(\boldsymbol{\varphi}_r|\boldsymbol{\varphi}_t, \boldsymbol{\theta}_i) + \mathrm{const}$~\cite{buckley_free_2017}. Beliefs are revised by natural-gradient descent~\cite{amari1998},
\begin{equation}
\boldsymbol{\theta}_{i+1} = \boldsymbol{\theta}_i - \boldsymbol{\eta}\,\mathcal{I}(\boldsymbol{\theta}_i)^{-1}\nabla_{\boldsymbol{\theta}}\mathcal{F}_i,
\end{equation}
where $\boldsymbol{\eta} = \mathrm{diag}(\eta_L, \eta_P, \eta_w)$ collects the learning rates and $\mathcal{I}$ is the Fisher information. For a von Mises concentration, $\mathcal{I}(\kappa) = A'(\kappa) = 1 - A(\kappa)^2 - A(\kappa)/\kappa$, with $A(\kappa) = I_1(\kappa)/I_0(\kappa)$ the Bessel ratio, i.e., the expected value of $\cos(x-\mu)$. For the hemifield weight, parameterised through its logit $\psi = \log\frac{w}{1-w}$, $\mathcal{I}(\psi) = w(1-w)$; in the update, $\boldsymbol{\theta}$ therefore denotes $(\kappa_L, \kappa_P, \psi)$, with $w$ recovered through the logistic function, $w = \sigma(\psi)$. Because the likelihood factorises, the three updates are decoupled.

\subsubsection{Update equations}
Let $\mathcal{VM}_c = \mathcal{VM}(\beta_r;\,\beta_t,\tilde\kappa_{P,i})$ and $\mathcal{VM}_m = \mathcal{VM}(\beta_r;\,\pi-\beta_t,\tilde\kappa_{P,i})$ denote the correct- and mirror-hemifield components of the polar mixture evaluated at the response on trial $i$, with $\tilde\kappa_{P,i} = \kappa_{P,i}\cos^2(\alpha_t)$, and $p_i = w_i\mathcal{VM}_c + (1-w_i)\mathcal{VM}_m$ the polar mixture likelihood. The updates are
\begin{align}
\kappa_{P,i+1} &= \kappa_{P,i} + \frac{\eta_P\,\cos^2(\alpha_t)}{A'(\kappa_{P,i})}\bigl[r_i - A(\tilde\kappa_{P,i})\bigr],
\label{eq:app_kappaP}\\
\psi_{i+1} &= \psi_i + \eta_w\,\frac{\mathcal{VM}_c - \mathcal{VM}_m}{p_i}, \qquad w_{i+1} = \sigma(\psi_{i+1}),
\label{eq:app_w}\\
\kappa_{L,i+1} &= \kappa_{L,i} + \frac{\eta_L}{A'(\kappa_{L,i})}\bigl[\cos(\alpha_r-\alpha_t) - A(\kappa_{L,i})\bigr],
\label{eq:app_kappaL}
\end{align}
where $\sigma$ is the logistic function and
\begin{equation}
r_i = \frac{w_i\cos(\beta_r-\beta_t)\,\mathcal{VM}_c + (1-w_i)\cos(\beta_r-\pi+\beta_t)\,\mathcal{VM}_m}{p_i}
\end{equation}
is the responsibility-weighted cosine alignment between response and target. Each bracketed term is a prediction error: the observed alignment minus the alignment expected under the current belief, for $\kappa_P$ evaluated at the scaled concentration $\tilde\kappa_{P,i}$. The cone of confusion enters twice. At perception time the listener has no access to the true lateral angle, so the generative model above conditions on the inferred cone $\alpha_r$. At update time the visual feedback reveals the true direction, so $\mathcal{F}$ in Eq.~\ref{eq:app_F} is evaluated with the true cone and the $\kappa_P$ step in Eq.~\ref{eq:app_kappaP} is scaled by $\cos^2(\alpha_t)$, the same factor used in the likelihood that fits $\boldsymbol{\theta}_0$ and the learning rates. For $w$, the Fisher information cancels the logistic derivative exactly, and the logit parameterisation keeps $w \in (0,1)$ without projection.

\subsection{Parameter estimation}
\label{app:parest}
Initial beliefs $\boldsymbol{\theta}_0$ were estimated per subject by maximum likelihood from the last 100 trials of the baseline block (naive localisation without feedback~\cite{majdak2010sound}). The learning rates $(\eta_L, \eta_P, \eta_w)$ are the only free parameters of the learning model. They were fitted per subject by minimising the mean negative log-likelihood of the responses in the feedback block under the sequential updates of Eqs.~\ref{eq:app_kappaP}--\ref{eq:app_kappaL}, using Bayesian adaptive direct search~\cite{acerbi2017practical}. Because the likelihood factorises, $(\eta_P, \eta_w)$ and $\eta_L$ are estimated independently.

\subsection{Model comparison and robustness}
\label{sec:app_robust}
\emph{Held-out comparison.} To test whether all three learning rates are needed, they were fitted on the odd-numbered feedback trials and the summed log-likelihood was evaluated on the even-numbered trials, with beliefs updated sequentially over all trials in both cases (Table~\ref{tab:app_robust}, left). Relative to the full model, freezing $\kappa_P$ ($\eta_P = 0$) reduced the held-out log-likelihood by 167 nats in total and freezing $w$ ($\eta_w = 0$) by 82 nats. Hemifield learning was required in every subject (NH12 only marginally), and polar-precision learning in all but NH17, for which $\eta_P$ was indistinguishable from zero (parametric bootstrap, $B = 100$, 95\% interval reaching zero). Over the feedback block, $\kappa_P$ rose by 2.8 to 4.5 in the other four subjects (95\% intervals excluding zero) and by 0.3 in NH17, with the largest relative gain in the two subjects with the lowest baseline $\kappa_P$ (NH15, NH18). A first-half/second-half split gave the same ordering.

\emph{Robustness to the $\cos^2$ scaling.} The Laplace approximation with a flat prior affects only the curvature of $\mathcal{F}$, not the location of the update, so it cannot mask a change in $\kappa_P$. The $\cos^2(\alpha)$ scaling changes the size of polar learning but not its direction (Table~\ref{tab:app_robust}, right): removing it from likelihood, update and baseline fit lowers the median $\eta_P$ from 0.015 to 0.006, $\kappa_P$ still rises by 0.6 to 2.8 over the feedback block, and the unscaled model fits worse with the same number of parameters (summed in-sample $\Delta\mathrm{LL} = -53$ nats, although NH15 and NH17 favour it). The rise of $w$ and the necessity of hemifield learning are unaffected. The rise of $\kappa_P$ is therefore not conditional on the geometric weighting of lateral targets. The unscaled model, however, does not respect the problem geometry: the cone of confusion collapses to a point at lateral positions ($\alpha = \pm 90^\circ$), where the polar angle is undefined, and polar errors become progressively less informative about spectral decoding as $|\alpha|$ increases.

\begin{table}[h]
\centering
\caption{Left: held-out comparison, log-likelihood differences in nats relative to the full model ($\Delta\mathrm{LL} < 0$: reduced model predicts worse). Right: $\cos^2$ robustness, polar-precision learning rate $\eta_P$ with bootstrap 95\% interval, with and without the $\cos^2(\alpha)$ scaling, and $\Delta\kappa_P$, the change of $\kappa_P$ over the feedback block without scaling.}
\label{tab:app_robust}
\setlength{\tabcolsep}{3pt}
\small
\begin{tabular}{|lr|rr|llr|}
\hline
 & & \multicolumn{2}{c|}{Held-out comparison} & \multicolumn{3}{c|}{$\cos^2$ robustness} \\
\hline
Subject & $\kappa_{P,0}$ & $\Delta\mathrm{LL}_{\eta_P=0}$ & $\Delta\mathrm{LL}_{\eta_w=0}$ & $\eta_P$ (cos$^2$) & $\eta_P$ (no cos$^2$) & $\Delta\kappa_P$ \\
\hline
NH12 & 7.5 &  $-5.7$ &  $-1.1$ & 0.020 [0.015, 0.033] & 0.005 [0.002, 0.010] & $+1.9$ \\
NH15 & 2.4 & $-54.2$ &  $-6.5$ & 0.015 [0.013, 0.026] & 0.016 [0.013, 0.025] & $+2.8$ \\
NH16 & 4.3 & $-39.0$ & $-39.0$ & 0.008 [0.006, 0.012] & 0.006 [0.005, 0.010] & $+1.6$ \\
NH17 & 4.3 &    0.0  & $-30.2$ & 0.002 [0, 0.012]     & 0.006 [0, 0.016]     & $+0.6$ \\
NH18 & 0.9 & $-68.2$ &  $-5.0$ & 0.062 [0.038, 0.084] & 0.043 [0.034, 0.079] & $+2.4$ \\
\hline
\end{tabular}
\end{table}

\end{document}